\documentclass[
  preprint,          
  reprint,
  amsmath,amssymb,  
  aps,              
  prl,              
  floatfix          
]{revtex4-1}

\usepackage{graphicx}
\usepackage{dcolumn}
\usepackage{booktabs}
\usepackage{bm}

\usepackage{amsmath}    
\usepackage{graphicx}   
\usepackage{hyperref}   
\usepackage{cleveref}   

\usepackage[table,svgnames,dvipsnames]{xcolor}
\usepackage[mathlines]{lineno}
\usepackage[normalem]{ulem}

\providecommand{\jcap}{JCAP}
\providecommand{\mnras}{MNRAS}

\crefname{figure}{Fig.}{Figs.}

\begin{document}

\preprint{APS/123-QED}

\title{Non-minimally coupled gravity constraints from DESI DR2 data}

\author{Jiaming Pan}
\email{jiamingp@umich.edu}
\affiliation{Department of Physics and Leinweber Center for Theoretical Physics,\\
 University of Michigan, 450 Church St, Ann Arbor, MI 48109}
 
\author{Gen Ye}%
\email{Corresponding author: ye@lorentz.leidenuniv.nl}
\affiliation{Institute Lorentz, Leiden University, PO Box 9506, Leiden 2300 RA, The Netherlands}

\date{\today}

\begin{abstract}
It has been observed that the hint about dynamical dark energy in the DESI BAO observation might point to non-minimally coupled gravity. We report the first $3\sigma$ evidence for non‐minimal coupling in a model‐agnostic effective field theory
(EFT) approach. In a non‐parametric reconstruction approach, we detect a clear departure from the General Relativity expectation of non-minimal coupling based on a combined analysis of DESI DR2 BAO together with CMB and Type Ia supernova data. 
Also, it is found that current data can constrain up to the quadratic order $(n=2)$ if the EFT function representing non-minimal coupling is Taylor expanded as a general function of the dark energy fraction $\Omega_{\rm{DE}}$, i.e. $\Omega^{\text{EFT}}(a)=\sum_{i=0}^{n} c^{\rm{EFT}}_i \Omega^i_{\text{DE}}(a)$. Our findings constitute a detection of modified‐gravity effects and call for a more flexible parametrization of the EFT functions than commonly used ones in literature.
\end{abstract}

\maketitle

\section{\label{sec:intro}Introduction}

The cosmological constant cold dark matter ($\Lambda$CDM) model has established itself as the standard model of cosmology by consistently explaining the majority of cosmological observations with only six parameters, including the state-of-the-art cosmic microwave background (CMB) observation \cite{Louis:2025tst} which is one of the most precise measurements ever made in cosmology. Baryon acoustic oscillation (BAO) is another precise cosmological measurement that measures the comoving sound horizon scale in the clustering of galaxies and quasars, as well as Lyman-$\alpha$ forest. The sound horizon scale is set by the CMB physics near recombination billions of years before when BAO is observed. Therefore, consistency between CMB and BAO is a stringent test of a cosmological model. The DESI Y1 BAO observation, for the first time, revealed hint about inconsistency between BAO and CMB in $\Lambda$CDM \cite{DESI:2024mwx}. This has triggered heated discussion about the robustness of data and methodology used to arrive at this conclusion, e.g. \cite{Wang:2024pui,Liu:2024gfy,Patel:2024odo,Efstathiou:2024xcq,Efstathiou:2024dvn,Wang:2024rjd,Colgain:2024xqj,Huang:2024qno,Wang:2024hwd}. Recently, the DESI Y3 data validated previous results and further reported a $2.3\sigma$ tension between CMB and BAO assuming $\Lambda$CDM \cite{DESI:2025zpo,DESI:2025zgx}, which is likely an indication of new physics beyond $\Lambda$CDM, in particular dynamical dark energy (DE) \cite{DESI:2024aqx,DESI:2024kob,Ishak:2024jhs,DESI:2025kuo}.

Furthermore, the DESI finding highlights the possibility of DE crossing the phantom divide, i.e. its equation of state $w_{\rm DE}$ crossing $-1$. This is first noticed by applying model independent reconstruction techniques to DESI Y1 \cite{DESI:2024aqx,Mukherjee:2024ryz,Dinda:2024ktd,Jiang:2024xnu,Gao:2025ozb,Johnson:2025blf,Yang:2025kgc,Berti:2025phi,DESI:2025kuo,Ormondroyd:2025exu,Ormondroyd:2025iaf,Yang:2024kdo,Ye:2024ywg,Berti:2025phi,Ormondroyd:2025exu,Pang:2024qyh} and further strengthened by the recent DESI Y3 data with both parametric and non-parametric methods \cite{DESI:2025kuo,Ormondroyd:2025iaf}. For a single non-interacting DE component, the existence of phantom crossing implies violation of the null energy condition in DE which not only rules out quintessence (a single minimally coupled canonical scalar field) as a DE candidate, see also \cite{Wolf:2024eph,Shlivko:2024llw,Bhattacharya:2024hep,Bhattacharya:2024kxp,Borghetto:2025jrk,Yin:2024hba} for some recent discussion, but also presents general difficulty to theory construction due to theoretical instabilities \cite{Carroll:2003st,Cline:2003gs,Dubovsky:2005xd,Creminelli:2008wc}. In principle, non-pathological phantom crossing can be realized by considering multi-fields \cite{Hu:2004kh,Feng:2004ad,Guo:2004fq,Wei:2005nw,Caldwell:2005ai,Cai:2007zv}, interactions in the dark sector \cite{Amendola:1999er,Das:2005yj,Khoury:2025txd,Chakraborty:2025syu,Li:2024qso} and modified gravity (MG) \cite{Carvalho:2004ty,Hu:2007nk,Nesseris:2006jc,Deffayet:2010qz,Chudaykin:2024gol,Ye:2024ywg,Wolf:2024stt,Yang:2025kgc,Yang:2025kgc}.

Important advance has been made to this end which reveals that observation might hint a specific form of MG, non-minimal coupling between gravity and matter \cite{Ye:2024ywg,Ye:2024zpk}, see also \cite{Wolf:2024stt,Wolf:2025jlc}. However, the hint is not conclusive given the data at that time.

\textit{In this paper we found the evidence of non-minimally coupled gravity in the latest observations reaches $3\sigma$ using model-independent analysis.} Specifically, we model the DE/MG effect relevant to cosmological observations using the effective field theory (EFT) framework. EFT is directly constructed from the assumed symmetry of the system (time-dependent spatial diffeomorphism invariance for cosmology) thus describes all possible theories satisfying said symmetry while being agnostic about any specific theory. Another possible source of bias is assuming a specific parametrization for the EFT function. To this end we adopt the non-parametric approach to perform a data-driven reconstruction of the non-minimal coupling function as a free function of time. As an example, our finding indicates that the commonly adopted EFT function parametrization $f_{\rm{EFT}}(a)=c_{\rm{EFT}}\Omega_{\rm{DE}}(a)$ \cite{Pujolas:2011he,Barreira:2014jha,Noller:2018wyv}, $\Omega_{\rm{DE}}(a)$ being the effective DE energy fraction, is indeed insufficient to capture the constrained information in the latest data and a more flexible parametrization $\Omega^{\rm{EFT}}(a)=\sum_{n\ge0}c^{\rm{EFT}}_n\Omega_{\rm{DE}}(a)$ is required.

\section{Methodology and data}

As pointed out by DESI \cite{DESI:2025kuo}, $\Lambda$CDM is in $3.1\sigma$ tension with the new BAO data combined with CMB. \cite{Ormondroyd:2025exu,Ormondroyd:2025iaf} also reports a mild discrepancy between DESI Y3 BAO and some of the SNIa data in $\Lambda$CDM. To avoid inconsistently combining datasets in tension, we thus need a background DE description compatible with observations while keeping agnostic about theoretical modeling. To this end we choose the Chevallier-Polarski-Linder (CPL) parametrization \cite{Chevallier:2001,Linder2003}
\begin{equation} \label{eq:cpl}
     w_{\text{DE}}(a) = w_0 + w_a(1-a),
\end{equation}
with the perturbation approximated in the post-Friedmann framework (PPF) \cite{Hu:2007pj}. This is a simple but good enough fit to current data \cite{DESI:2025kuo,Ormondroyd:2025iaf}. The other species follow their standard description in $\Lambda$CDM. The Universe is assumed to be spatially flat. The resultant background cosmological model will be referred to as $w_0w_a$CDM as usual. In particular, neutrinos are treated as two massless and one massive with mass $0.06$ eV following Planck \cite{Planck:2018vyg}, which has been confirmed to be consistent with the new DESI BAO in $w_0w_a$CDM \cite{DESI:2025ejh}. We adopt uninformative uniform priors for all cosmological parameters as reported in Table.\ref{tab:priors}.

The DE background and linear perturbation dynamics are described in the EFT framework. The quadratic EFT of DE with at most second order equation of motion is \cite{Bloomfield:2012ff,Gubitosi:2012hu}
\begin{widetext}
\begin{equation} \label{eq:eft}
\begin{split}
    S = \int d^4x\,\sqrt{-g}\Biggl\{ &\frac{M_p^2}{2}\Bigl[ 1 + \Omega^{\text{EFT}}(\tau) \Bigr] R + \Lambda(\tau) - c(\tau)\,a^2\,\delta g^{00} \\
    &+ \frac{M_p^2 H_0^2}{2}\,\gamma_1(\tau) \left( a^2\,\delta g^{00} \right)^2
    -\frac{M_p^2 H_0}{2}\,\gamma_2(\tau)\,a^2\,\delta g^{00}\,\delta K^\mu{}_\mu\\[1mm]
    &-\frac{M_p^2}{2}\,\gamma_3(\tau)\Biggl[
    \left( \delta K^\mu{}_\mu \right)^2 - \delta K^\mu{}_\nu\,\delta K^\nu{}_\mu
    -\frac{a^2}{2}\,\delta g^{00}\,\delta R
    \Biggr]
    \Biggr\} + S_m\left[g_{\mu\nu},\chi_m\right]\,
\end{split}
\end{equation}
\end{widetext}
where we have adopted the notion in \cite{Hu:2014oga}. Eq.\eqref{eq:eft} covers the background and linear perturbation dynamics of all covariant scalar-tensor theories belonging to the Horndeski class \cite{Horndeski:1974wa}, with the mapping between the two described in \cite{Gleyzes:2013ooa,Bloomfield:2013efa,Kennedy:2017sof}. The lagrangian coefficients $\{\Omega^{\rm{EFT}}, \Lambda, \gamma_{1,2,3}\}$ of \eqref{eq:eft}, also refereed as the EFT functions, are free functions of time to be specified ($c(\tau)$ is not independent and is derived from $\{\Omega^{\rm{EFT}},\Lambda\}$). Non-minimal coupling is uniquely described by $\Omega^{\rm{EFT}}\ne0$ in \eqref{eq:eft}. We therefore focus on the non-minimal coupling EFT function $\Omega^{\text{EFT}}$ and set $\gamma_{1,2,3}=0$. As shown in \cite{Ye:2024ywg}, this is justified because $\gamma_{1,2,3}$ has little effect on stabilizing the theory. The only remaining EFT function $\Lambda(\tau)$ is fixed by choosing the $w_0w_a$CDM background. We use \texttt{EFTCAMB} \cite{Hu:2013twa,Raveri:2014cka}, based on \texttt{CAMB} \cite{Lewis:1999bs,Howlett:2012mh}, to compute cosmology, which fully implements \eqref{eq:eft} and the relevant theoretical stability criteria. We limit our scope to the late Universe thus only evolve DE perturbation and check for stability for $z<9$, while perturbations are described by general relativity (GR) for $z>9$. We perform Monte Carlo Markov chain (MCMC) analysis using the \texttt{Cobaya} sampler \cite{Torrado:2020dgo,2019ascl.soft10019T} to fit to data and derive posteriors of the cosmological and reconstruction parameters. We use the Gelman-Rubin diagnostic $R-1<0.02$~\cite{Gelman:1992zz} as our convergence criteria.

\begin{table}[ht]
\centering
\renewcommand{\arraystretch}{1.2} 
\begin{tabular}{lcc}
\hline
Parameter                  & Prior                       & Default \\
\hline
$\omega_b$                 & $\mathcal{U}[0.005,0.1]$     & --      \\
$\omega_c$                 & $\mathcal{U}[0.001,0.99]$    & --      \\
$H_0$ [km/s/Mpc]           & $\mathcal{U}[20,100]$       & --      \\
$n_s$                    & $\mathcal{U}[0.8,1.2]$       & --      \\
$\ln(10^{10}A_s)$          & $\mathcal{U}[1.61,3.91]$     & --      \\
$\tau$                   & $\mathcal{U}[0.01,0.8]$      & --      \\
\hline
$w_0$                    & $\mathcal{U}[-3,1]$         & $-1$    \\
$w_a$                    & $\mathcal{U}[-3,2]$         & $0$     \\
\hline
$\Omega^{\rm{EFT}}_{i}$    & $\mathcal{U}[-1,1]$         & $0$     \\
$c^{\rm{EFT}}_{i}$         & $\mathcal{U}[-5,5]$         & $0$     \\
\hline 
\end{tabular}
\caption{Summary of the uniform priors on the MCMC parameters. ``Default'' refers to the $\Lambda$CDM value of the corresponding parameter.}
\label{tab:priors}
\end{table}

\subsection{Non-parametric reconstruction of $\Omega^{\rm{EFT}}(a)$}

Data-driven reconstruction techniques are widely utilized in the study of MG, where prior preference for a particular theoretical model is lacking \cite{Zhao:2009fn,Zhao:2010dz,Hojjati:2011xd,Silvestri:2013ne,Raveri:2017qvt,Espejo:2018hxa,Raveri:2021dbu,Pogosian:2021mcs,deBoe:2024gpf,Zhao:2017cud,Zhao:2012aw,Raveri:2019mxg}. In this paper, we perform a the non-parametric reconstruction of the non-minimal coupling function $\Omega^{\rm{EFT}}(a)$ by binning its function values $\Omega^{\rm{EFT}}_i\equiv\Omega^{\rm{EFT}}(a_i)$ at six uniformly spaced scale factors $a_i=[0.5, 0.6, 0.7, 0.8, 0.9, 1.0]$, corresponding to $z=[1, 0.67, 0.43, 0.25, 0.11, 0]$. The value of $\Omega^{\rm{EFT}}(a)$ is obtained by interpolation between the nodes for $1>a>0.5$. Since our focus is on the DE and phantom crossing behavior at $z<1$, the value of $\Omega^{\rm{EFT}}(a)$ for $0<a<0.5$ is determined by smooth extrapolation of the $1>a>0.5$ region to the mean value of nodes $\Bar{\Omega}^{\rm{EFT}}\equiv\frac{1}{6}\sum_{i=1}^{6}\Omega^{\rm{EFT}}_i$. In the reconstruction, the node values $\{\Omega^{\rm{EFT}}_i\}$ are sampled together with all the other parameters in MCMC with uninformative uniform priors reported in Table.\ref{tab:priors}. This non-parametric reconstruction provides an independent view on the preferred functional forms of \(\Omega^{\text{EFT}}(a)\). Its results can be used to determine which parametrization form is suitable for the data considered.

\subsection{Parametric reconstruction of $\Omega^{\rm{EFT}}(a)$}
DE remains subdominant in the matter dominant era and only becomes relevant when the Universe starts accelerated expansion. Therefore, a natural assumption is that the non-minimal coupling effect associated with the DE component, characterized by $\Omega^{\rm{EFT}}$, also follows a similar evolution, i.e. close to GR early on and got triggered when DE becomes relevant. We therefore propose the following parametrization
\begin{equation}
    \Omega^{\text{EFT}}(a)=\sum_{i=0}^{n} c^{\rm{EFT}}_i \Omega^i_{\text{DE}}(a).
\label{eq:omega_unified}
\end{equation}
with the effective energy fraction of DE defined as
\begin{equation}\label{eq:Ode}
    \Omega_{\rm{DE}}(a)\equiv 1 - \frac{\rho_{\rm{matter}}(a)}{3M_p^2H^2(a)}.
\end{equation}
$\rho_{\rm{matter}}$ is the total energy density of all species except for the DE component. If only $c_1\ne0$, eq.\eqref{eq:omega_unified} reduces to the $f_{\rm{EFT}}=c_{\rm{EFT}}\Omega_{\rm{DE}}$ parametrization widely adopted in the literature. However, as we will show in the next section, this linear parametrization is no longer sufficient and one needs to consider up to $n=2$ to capture the relevant information in the latest observational data \footnote{We found current data has difficulty constraining $n=3$.}. The prior for the Taylor coefficients $\{c^{\rm{EFT}}_i\}$ is reported in Table.\ref{tab:priors}.

\subsection{Dataset}

We consider the following datasets in this paper:
\begin{itemize}
    \item \textbf{DESI BAO:} The full DESI DR2 BAO measurement \cite{DESI:2025zpo,DESI:2025zgx}. We note that BAO analysis has been shown to be robust to modified gravity models \cite{Pan2024JCAP}.
    \item \textbf{DESY5 SN:} The DESY5 sample containing 1635 DES SNIa in the redshift range $0.1<z<1.13$ combined with 194 external low-redshift SNIa in $0.025<z<0.1$  \cite{DES:2024jxu}.
    \item \textbf{CMB:} 

    The \texttt{CamSpec} version of the Planck PR4 high-$\ell$ TTTEEE \cite{Efstathiou:2019mdh} and Planck PR3 low-$\ell$ TTEE \cite{Planck:2019nip} data, along with Planck PR4 lensing \cite{Carron:2022eyg}. 

\end{itemize}

There have been discussion that the SNIa systematics and the difference in excess lensing-like smoothing in CMB temperature might impact the constraint on DE, see e.g. \cite{Peng:2025nez,Ishak:2024jhs,Efstathiou:2024xcq,Huang:2025som}. We have checked that switching the SNIa data to Panthon+ \cite{Brout:2022vxf} or the CMB TTTEEE data to \texttt{LoLLiPop} and \texttt{HiLLiPop} version of the Planck PR4 likelihood \cite{Tristram:2023haj} has only marginal impact on the posteriors and does not change the conclusion.

\begin{figure}[h]
\includegraphics[width=1\linewidth]{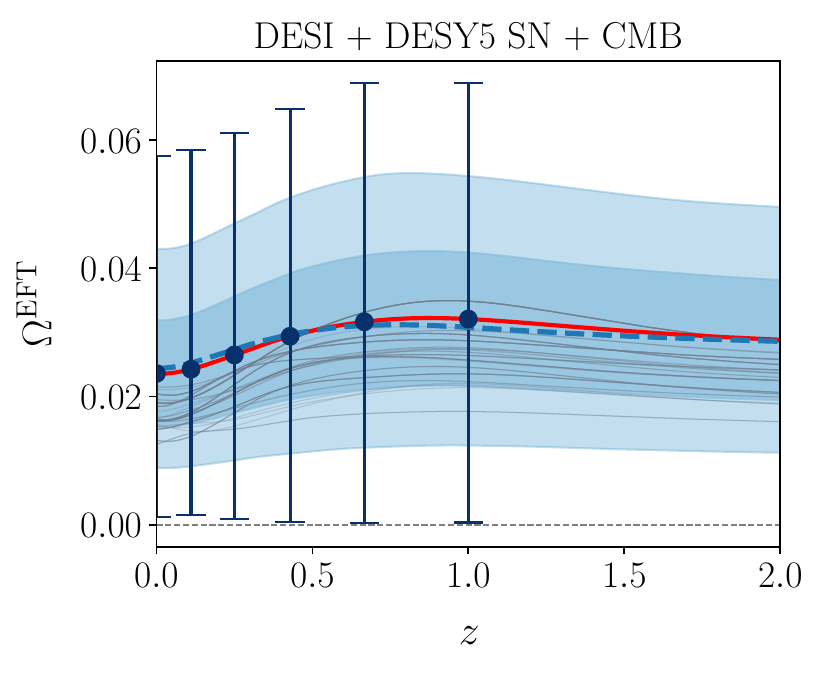}
\caption{Reconstruction of the non-minimal coupling $\Omega^{\mathrm{EFT}}(a)$. The red line shows the mean reconstruction and the blue bands represent the 68\% and 95\% confidence intervals. Blue points with 3$\sigma$ error bars indicate the node values $\{\Omega^{\rm{EFT}}_i\}$ directly constrained by the data. The solid grey lines are individual realizations that illustrate the preferred functional forms. In the plotted redshift range of interest $0<z<2$, the reconstructed $\Omega^{\rm{EFT}}$ deviates from its GR value ($\Omega^{\mathrm{EFT}}(a)=0$, shown as a horizontal dashed line) by $\sim3\sigma$. Dashed line indicates the first mode in principal component analysis, which has the smallest uncertainty. }
\label{fig:Omega_recon}
\end{figure}

\section{Results}

\cref{fig:Omega_recon} shows the reconstruction of $\Omega^{\text{EFT}}(a)$ using the combination of DESI, DESY5 SN, and CMB. The mean values as a function of redshfits are shown in the solid red lines, with different blue bands indicate 68$\%$ and 95$\%$ confidence intervals. The blue data points are the mean value and $3\sigma$ posterior of the six $\{\Omega^{\rm {EFT}}_i\}$ nodes, which are directly sampled and constrained by the data in the MCMC. The solid grey lines are 60 realizations of the reconstruction results as a illustration of the preferred shapes of $\Omega^{\text{EFT}}(a)$. Notably, the reconstructed values of the non-minimal coupling function $\Omega^{\rm{EFT}}$ deviates from the GR prediction (i.e., \(\Omega^{\text{EFT}}(a)=0\)) at $\sim3\sigma$ in the plotted range $0<z<2$. The shape constraint is consistent with previous study \cite{Ye:2024ywg} but shows a considerably higher significance of non-minimal coupling ($\Omega^{\rm{EFT}}\ne0$) thanks to the improved observation data. This is also consistent with the finding of \cite{Frusciante:2019xia} in that stability generally prefers $\Omega^{\rm{EFT}}>0$. 

To quantify the degree to which the reconstruction results from the GR expectation, we apply principal component analysis (e.g.\ \cite{Huterer:2002hy,Huterer:2004ch,Crittenden:2005wj}) directly on the MCMC samples of the non-minimal coupling function \(\Omega^{\mathrm{EFT}}(a)\). The first principal component—i.e.\ the mode with the smallest uncertainty—deviates from the GR prediction \(\Omega^{\mathrm{EFT}}(a)=0\) at the \(2.9\,\sigma\) level.  In Fig.~\ref{fig:Omega_recon}, this leading mode is plotted as a blue dashed line, closely tracking the mean of reconstruction results, highlighting its dominant contribution to the overall preference for \(\Omega^{\mathrm{EFT}}>0\). Accounting for both the leading and subleading principal components, we detect
$\Omega^{\mathrm{EFT}}\neq 0$ at $\gtrsim3\sigma$ significance.

\begin{figure}[h]
\includegraphics[width=\linewidth]{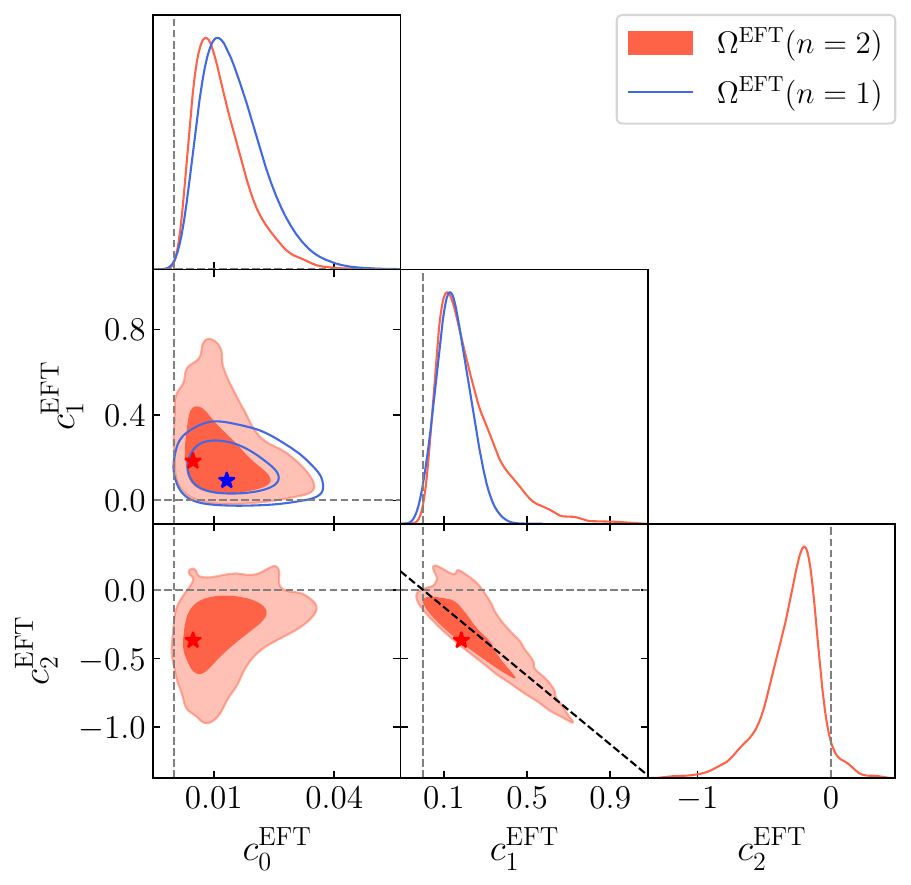}
\caption{68\% and 95\% posterior constraints on the coefficients of non-minimal coupling parameterization $
\Omega^{\rm EFT}(a)=\sum_{n\ge0} c_n^{\rm EFT}\Omega_{\rm DE}^n(a)
$. The blue contour corresponds to the first-order expansion $\Omega^{\rm{EFT}}(n=1)$, and the red contour represents the second-order expansion $\Omega^{\rm{EFT}}(n=2)$. The star marks the MAP point for each of the model. Both MAP points fall in the corresponding $1\sigma$ contour, suggesting that projection effect is small. The black dotted line plots the expected correlation direction $c_1^{\rm{EFT}}/c_2^{\rm{EFT}}=-0.8$, corresponding to a peak in $\Omega^{\rm{EFT}}$ near $z=0.6$.}
\label{fig:Contours}
\end{figure}

\begin{figure*}[]
\includegraphics[width=1\linewidth]{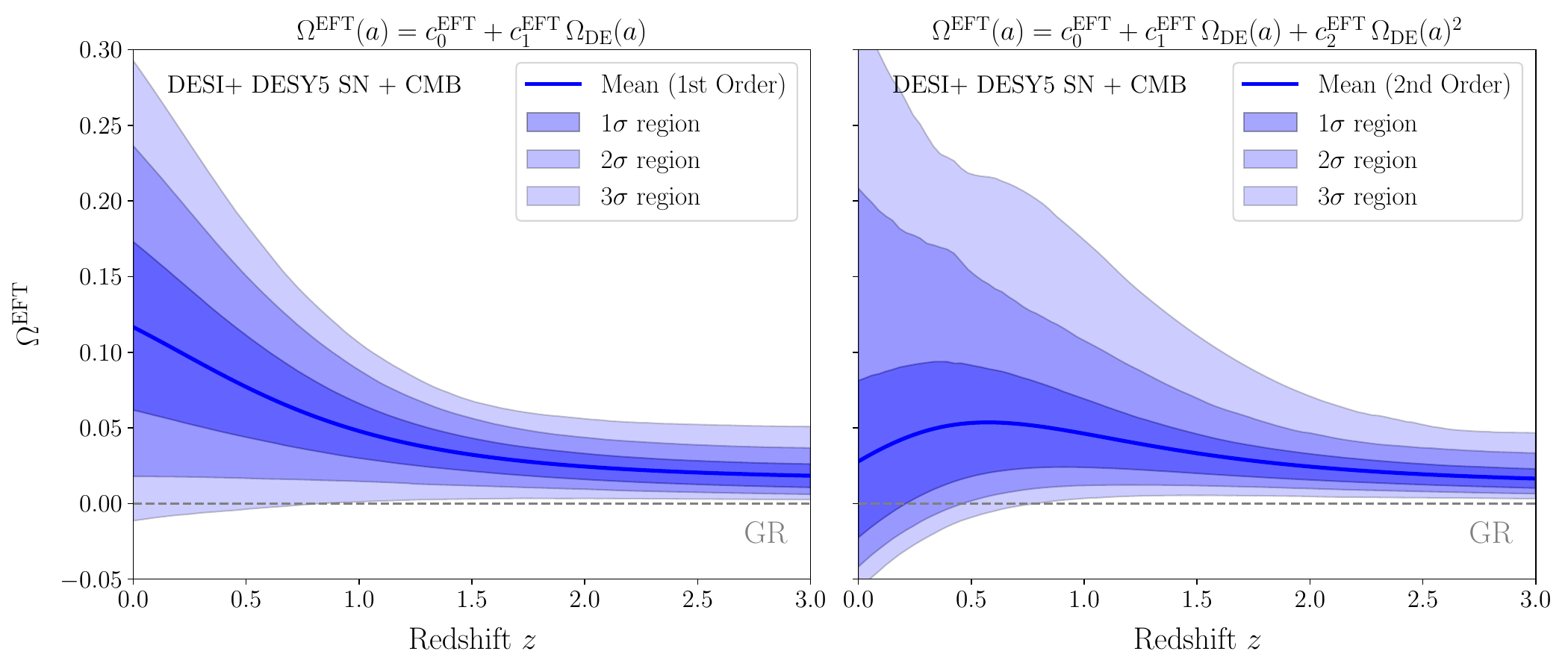}
\caption{Left panel: First-order expansion of $\Omega^{\rm EFT}(a)$ parameterization constrained by DESI, DESY5 SN, and CMB data. Right panel: Second-order expansion. Both parameterizations indicate larger than $2\sigma$ signal for the non-minimal coupling. The second-order expansion achieves a better fit than the first-order expansion, with a more similar shape to the reconstruction results.}
\label{fig:Omega_Parametric}
\end{figure*}

Next, we explore the parameterization in Eq.~\eqref{eq:omega_unified}, which introduces fewer free parameters compared to non-parametric approach. \cref{fig:Contours} plots the posterior constraints on the parametrization coefficients and \cref{fig:Omega_Parametric} shows the constraints on the non-minimal coupling terms derived using the parametric approach, constrained by DESI, DESY5 SN, and CMB data. The left panel displays the first-order parametrization ($n=1$ in eq.\eqref{eq:omega_unified}) of $\Omega^{\rm EFT}(a)$ as a function of redshift, and the right panel shows the second-order parametrization ($n=2$). The second-order parameterization provides a more accurate fit than the first-order one, successfully capturing the bump in the reconstruction results at $z\sim0.6$ (\cref{fig:Omega_recon}). This improvement is also reflected in the improvement of fit, quantified by $\Delta\chi^2$ value relative to $\Lambda$CDM summarized in Table~\ref{tab:deltaChi2}. We note that both parameterizations indicate a clear detection for non-minimal coupling by more than $2\sigma$ as illustrated in \cref{fig:Contours}.

In \cref{fig:Contours}, we present the constraints on the non-minimal coupling parameterization coefficients from the combined DESI, DESY5 SN, and CMB datasets. The blue contour corresponds to the first-order ($n=1$) parameterization in Eq.~\eqref{eq:omega_unified}, and the red contour is the second-order ($n=2$) result. We obtained the following $1\sigma$ posterior constraints for the non-minimal coupling parameterization coefficients:

\begin{equation}
\left.
\begin{array}{rcl}
c_0^{\rm EFT} &=& 0.0151^{+0.0051}_{-0.0096},\\[0.2cm]
c_1^{\rm EFT} &=& 0.149^{+0.072}_{-0.090}
\end{array}
\right\} \quad \text{}\Omega^{\text{EFT}}(n=1)\text{}
\label{eq:OmegaEFT_n1}
\end{equation}

\begin{equation}
\left.
\begin{array}{rcl}
c_0^{\rm EFT} &=& 0.0118^{+0.0034}_{-0.0082},\\[0.2cm]
c_1^{\rm EFT} &=& 0.222^{+0.063}_{-0.18},\\[0.2cm]
c_2^{\rm EFT} &=& -0.30^{+0.21}_{-0.13}
\end{array}
\right\} \quad \text{}\Omega^{\text{EFT}}(n=2)\text{}
\label{eq:OmegaEFT_n2}
\end{equation}
The coefficient \(c_0^{\rm EFT}\) deviates by more than \(2\sigma\) from the GR prediction in both parameterizations, indicating a consistent signal for non-minimal coupling of gravity. The higher order terms coefficients (\(c_1^{\rm EFT}\) and \(c_2^{\rm EFT}\)) deviate from GR predictions at more than 1$\sigma$. 
The strong correlation between $c^{\rm{EFT}}_1$ and $c^{\rm{EFT}}_2$ in \cref{fig:Contours} is what captures the bump feature near $z=0.6$ in the reconstruction in \cref{fig:Omega_recon}, because fixing the peak location to $z\sim0.6$ in the $n=2$ case requires $c_1^{\rm{EFT}}/c_2^{\rm{EFT}}=-2\Omega_{\rm{DE}}|_{z=0.6}$. Assuming $\Lambda$CDM and setting $\Omega_m=0.3$, one has $\Omega_{\rm{DE}}(z=0.6)\simeq0.4$ and the estimated correlation direction $c_1^{\rm{EFT}}/c_2^{\rm{EFT}}\simeq-0.8$, consistent with the numerical results in \cref{fig:Contours}.  

 Notably, the second-order expansion not only offers a better fit—evidenced by an improved \(\Delta\chi^2\) relative to \(\Lambda\)CDM—but also captures additional features seen in the non-parametric reconstruction. This implies that some commonly used EFT function parameterization, e.g. $f_{\rm{EFT}}(a)=c_{\rm{EFT}}\Omega_{\rm{DE}}(a)$ as a simplified version of the first-order expansion, in the literature may be overly restrictive for effectively constraining the EFT parameters with latest observations, underscoring the need for more flexible functional forms in future analyses. It is also the reason why the scaling parametrization $\Omega^{\rm{EFT}}\propto a^s$ failed to capture the non-minimal coupling effect in \cite{Ishak:2024jhs}. Furthermore, the EFT functions in this study can be converted into the EFT functions in the \(\alpha\)-basis \cite{Bellini2014}. When parameterized appropriately based on our study's results, these MG functions should also capture the non-minimal coupling signal.

Projection effects—where the mean of the marginalised posterior can be offset from the maximum a posteriori (MAP) value due to weak statistical constraints which is likely to happen when constraining modified gravity models (see \cite{DESI:2024jis} for further details about projection effects)—are small in this analysis. In particular, the MAP values, indicated by the blue and red star markers for the first- and second-order expansions parameterizations respectively, lie within \(1\sigma\) of the statistical uncertainties \footnote{The MAP values are obtained from \texttt{prospect} \cite{prospect}.}. This demonstrates that any bias from projection effects is small, thereby reinforcing the robustness of the non-minimal coupling signal inferred from the DESI DR2 CMB, and SNIa datasets.

\begin{figure}[t]
  \centering
  \includegraphics[width=\linewidth]{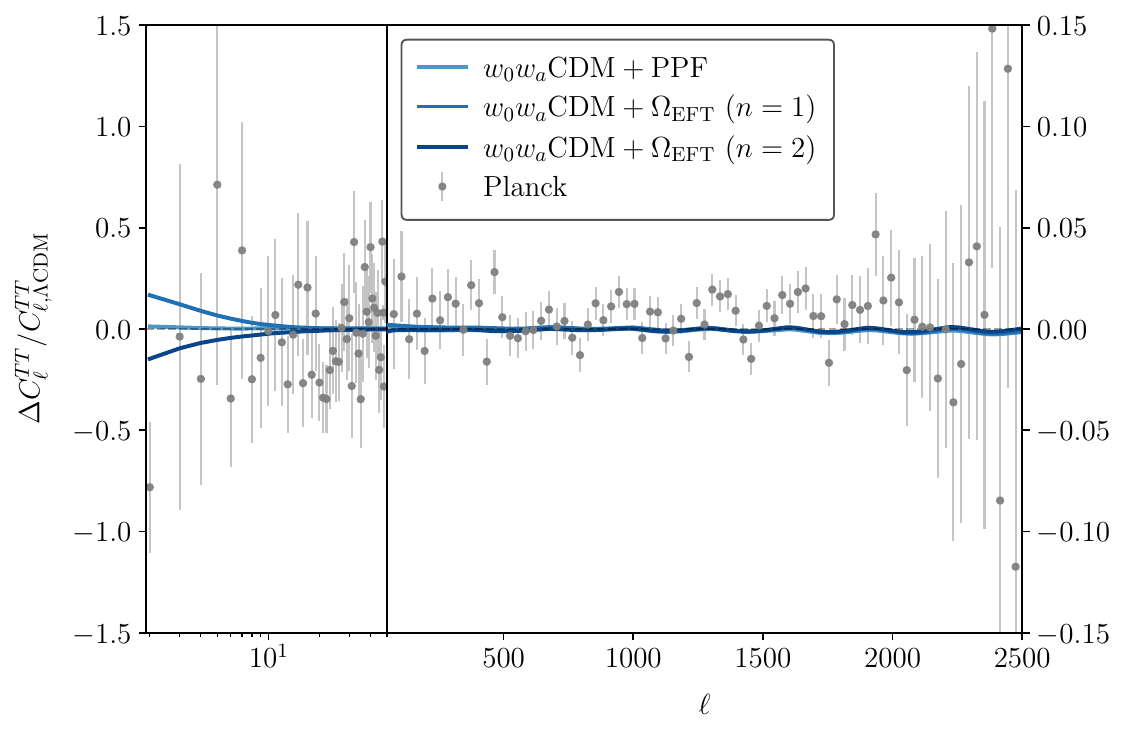}
  \caption{\label{fig:TT_residuals}
  Impact on the CMB temperature power spectrum shown as fractional residuals relative to the best–fit $\Lambda$CDM spectrum, using the combined datasets DESI DR2 BAO, DESY5 SN, and Planck2018 CMB. Black points with error bars show the binned Planck TT measurements; colored curves are the MAP predictions of the extended late–time models discussed in the text, each shown relative to $\Lambda$CDM.
  }
\end{figure}

To show the effects of late–time MG perturbations in the CMB spectra, we plot the CMB temperature residuals 
\[
\frac{\Delta C_\ell^{TT}}{C_{\ell,\Lambda{\rm CDM}}^{TT}} \equiv 
\frac{C_\ell^{TT,{\rm model}} - C_\ell^{TT,\Lambda{\rm CDM}}}
     {C_{\ell,\Lambda{\rm CDM}}^{TT}}
\]
in \cref{fig:TT_residuals}.
The black points show the binned Planck 2018 TT data \cite{Planck:2019nip}, while the colored curves are the predictions corresponding to the MAP points of $w_0w_a$CDM+PPF and two non–minimal coupling parameterizations, each shown relative to the best–fit $\Lambda$CDM spectrum.
Since only late-time MG is considered, the main impact of non-minimal coupling is deviation from $\Lambda$CDM prediction caused by the late ISW effect in the low $\ell$ range, where data is dominated by cosmic variance. To this end, observation of late time perturbation evolution, such as weak lensing and redshift space distortion, is needed to further constrain the model.

\begin{table}[ht]
\centering
\begin{tabular}{lc}
\hline
Parametrization      & $\Delta\chi^2$ ($\Lambda$CDM) \\
\hline
$w_0w_a$CDM + PPF                        & -20.5                \\
$w_0w_a$CDM + $\Omega^{\text{EFT}}(n=1)$ & -20.6                \\
$w_0w_a$CDM + $\Omega^{\text{EFT}}(n=2)$ & -22.3                \\

\hline
\end{tabular}
\caption{Comparison of \(\Delta\chi^2\) values for non-minimal coupling parametrizations, based on the DESI + DESY5 SN + CMB datasets.}
\label{tab:deltaChi2}
\end{table}

\section{Conclusion}

We have presented a model‑agnostic analysis of non-minimal coupling of dark energy using a combination of DESI DR2 BAO, DESY5 SN, and CMB datasets.
Our analysis reveals a $3\sigma$ evidence of a specific modified gravity effect, \textit{non-minimal coupling between gravity and matter}, in the combined datasets. The indication for non-minimally coupled gravity derives from requiring the stability of the theory, in the EFTofDE framework, that supports the observed evolution. This signal is found using a non-parametric reconstruction method modeling dark energy as a single canonical scalar field. Furthermore, we introduce an improved, flexible parametrization eq.\eqref{eq:omega_unified} of the EFT functions which indicates more than \(2\sigma\) non-minimal coupling signal. This extension of $\Lambda$CDM fits combined dataset as good as $w_{0}w_{a}$CDM with additional non-minimal coupling terms. Our findings challenge the commonly used $f_{\rm{EFT}}(a)=c_{\rm{EFT}}\Omega_{\rm{DE}}(a)$ parametrization in the literature which would overlook the MG signals. These findings remain robust against switching the SNIa data to Pantheon+ \cite{Brout:2022vxf} or the CMB likelihood to \texttt{LoLLiPop} and \texttt{HiLLiPop} version of the Planck PR4 likelihood \cite{Tristram:2023haj}.

Finally, we stress that both SNIa and BAO are geometric probes measuring the background cosmological evolution and MG effect has been turned off for $z>9$ in this study to focus on late-time DE effects. Since we have assumed the same $w_0w_a$ background, inclusion of the non-minimal coupling function $\Omega_{\rm EFT}$ does not significantly improve fit to SNIa, BAO and CMB, which mainly probes perturbation dynamics in the early times, see Table.\ref{tab:deltaChi2}. In fact, the signal for non-minimal coupling of gravity arises from requiring a theoretically stable phantom crossing cosmology, as indicated by the data, within the EFT framework. Therefore, to further assess the evidence for MG, one needs additional information on late-time perturbation evolution, from e.g. weak lensing surveys \cite{Euclid:2024yrr,LSST:2008ijt} and full shape galaxy clustering \cite{DESI:2024jis}, and consistently take into account its effect in the early Universe, e.g. \cite{Wang:2024dka,Ye:2024zpk,Pan:2025upl}.

\textbf{Acknowledgments:} 
JP acknowledges support from the Leinweber Center for Theoretical Physics and DOE under contract DE-SC009193, and also thanks Otávio Alves for helpful discussions. GY acknowledges support by NWO and the Dutch Ministry of Education, Culture and Science (OCW) (grant VI.Vidi.192.069). This research was supported in part through computational resources and services provided by Advanced Research Computing at the University of Michigan, Ann Arbor. 

\bibliography{reference}

\end{document}